\documentclass[sigconf]{acmart}

\usepackage{algpseudocode}
\usepackage{algorithm}
\usepackage{multirow}
\usepackage{amssymb}
\usepackage{dsfont}
\usepackage{animate}

\newcommand*{\vertbar}{\rule[-1ex]{0.5pt}{2.5ex}}

\copyrightyear{2021}
\acmYear{2021}
\setcopyright{acmlicensed}\acmConference[MM '21]{Proceedings of the 29th ACM International Conference on Multimedia}{October 20--24, 2021}{Virtual Event, China}
\acmBooktitle{Proceedings of the 29th ACM International Conference on Multimedia (MM '21), October 20--24, 2021, Virtual Event, China}
\acmPrice{15.00}
\acmDOI{10.1145/3474085.3475180}
\acmISBN{978-1-4503-8651-7/21/10}

\begin{document}
\fancyhead{}
\title{Dual Learning Music Composition and Dance Choreography}

\author{Shuang Wu}
\email{wushuang@outlook.sg}
\affiliation{%
  \institution{Nanyang Technological University}
  \country{}
}

\author{Zhenguang Liu}
\authornote{Corresponding Authors}
\email{liuzhenguang2008@gmail.com}
\affiliation{%
  \institution{Zhejiang Gongshang University}
  \country{}
}

\author{Shijian Lu}
\email{Shijian.Lu@ntu.edu.sg}
\affiliation{%
  \institution{Nanyang Technological University}
  \country{}
}

\author{Li Cheng}
\authornotemark[1]
\email{lcheng5@ualberta.ca}
\affiliation{%
  \institution{University of Alberta}
  \country{}
}


\begin{abstract}
Music and dance have always co-existed as pillars of human activities, contributing immensely to the cultural, social, and entertainment functions in virtually all societies. Notwithstanding the gradual systematization of music and dance into two independent disciplines, their intimate connection is undeniable and one art-form often appears incomplete without the other. Recent research works have studied generative models for dance sequences conditioned on music. The dual task of composing music for given dances, however, has been largely overlooked. In this paper, we propose a novel extension, where we jointly model both tasks in a dual learning approach. To leverage the duality of the two modalities, we introduce an optimal transport objective to align feature embeddings, as well as a cycle consistency loss to foster overall consistency. Experimental results demonstrate that our dual learning framework improves individual task performance, delivering generated music compositions and dance choreographs that are realistic and faithful to the conditioned inputs.
\end{abstract}


\begin{teaserfigure}
\centering
\includegraphics[width=0.88\textwidth]{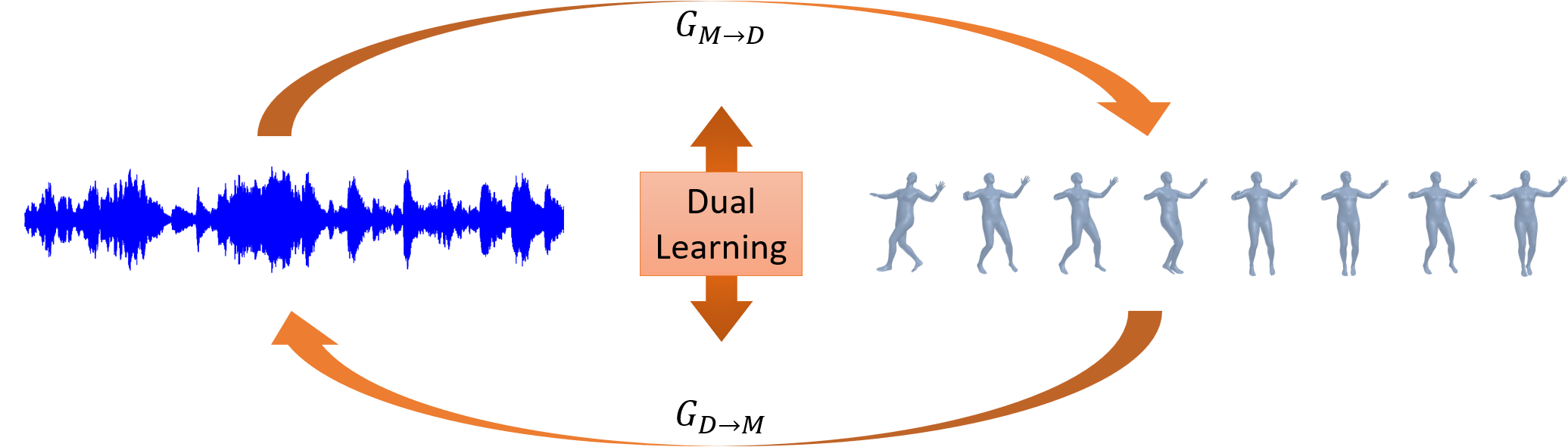}
\caption{We propose a novel problem where we concurrently learn dance choreography and music composition. Specifically, our framework consists of two networks, $G_{M\to D}$ for generating 3D dance choreographs from input music, and $G_{D\to M}$ that synthesizes music compositions given dance sequences. We leverage the duality of these tasks to extract the common underlying themes and ensure consistency between the generated output and the conditional input.}
\label{fig:task}
\end{teaserfigure}

\maketitle

\section{Introduction}
From an evolutionary perspective, music and dance have played a vital role for the social function of the human species \cite{wang2015hypothesis}. They are ubiquitous in human activities, ranging from personal entertainment to social functions and ceremonial activities. On top of forming an indispensable tapestry in human culture, they are also integral to modern civilization and contribute immensely to our individual and social well-being. Over the past centuries, music and dance are gradually systematized into two separate art-forms but their intimate and deep connection is unmistakable. Both entail expressing our internal emotions as external movements. For dance, the medium of expression is visual in the form of body movements whereas for music, movements manifest auditorily through melodies and rhythms.

It is widely acknowledged that music evokes motions and feelings of intentionality \cite{small1998musicking}. We may find ourselves moving along to the beats and dancing to music, perhaps even unaware of our movements. Neuroscience research elucidates how music and dance activities may involve similar stimuli in our brain \cite{brown2008neuroscience,levitin2009current}. Recently, the artificial intelligence community has also taken an interest. Several works \cite{fan2011example,lee2013music,tang2018dance,lee2019dancing,ren2020self,huang2021dance} have investigated the task of generating dance choreographs from music.

Reciprocally, dance would appear incomplete and unadorned if there were no accompanying music. Since generating music from dance remains largely overlooked in the literature, we propose to concurrently tackle this dual task. As summarized in Fig. \ref{fig:task}, we develop a bi-directional generative model for synthesizing realistic and matching dance from music \emph{and} music from dance. This dual learning has the additional advantage of enhancing the modeling of each modality. Given the popularity of music and dance videos in entertainment and multimedia, our ultimate goal is to enable effective engagement of the public, contribute to the user experience, and benefit the vast community of musicians and dancers.

There are several challenges in this task.

\textbf{Cross-domain generation} \quad Translating between music and dance constitutes a cross-domain sequence-to-sequence generative modeling task. Most source-to-target domain learning tasks entail image-to-image \cite{isola2017image,yi2017dualgan,zhu2017unpaired,lin2018conditional} or language translation \cite{he2016dual}, for which the data lies in topologically identical spaces. However, our task is further complicated by the fact that the ambient spaces for our data have totally distinct topological properties. Specifically, music is represented as waveforms, whereas a dance sequence is represented as 3D motion trajectories on a pose manifold. This increases the difficulty of learning a network mapping between music and dance with realistic outputs. 

\textbf{Creativity and Diversity} \quad Earlier works in dance choreography \cite{fan2011example,lee2013music} adopted a similarity retrieval approach which simply glues together dance moves from a learned template and is lacking in both innovation and diversity. \cite{tang2018dance} utilized a sequence-to-sequence model with Long Short Term Memory (LSTM) units which suffered from the limitation of a single output. However, multiple interpretations for music composition or dance choreography is commonplace and diversity should also be reflected in the model. One viable approach \cite{lee2019dancing,ren2020self} is to employ Generative Adversarial Networks (GANs) that enables a distribution of plausible outputs instead of a single deterministic one. 

\textbf{Consistency between music and dance} \quad On top of securing the realism of the generated music pieces or dance sequences, we need to ensure harmony between the generated output and the conditioned input. In other words, we need to extract the shared themes and intentionality between the two media as domain-independent features and ensure that such domain-independent abstractions are reflected in the target output. Furthermore, the kinematic beats in dance and the acoustic beats in music should be aligned.

To address these challenges, our proposed approach employs transformer network architectures \cite{vaswani2017attention} as encoders and decoders in a sequence-to-sequence framework. We incorporate a full attention mechanism \cite{devlin2018bert} for feature learning in both the music and dance domain. This has the key advantage of a global level understanding of underlying themes. We instill diversity into our model by concatenating a random vector with the encoded feature. Furthermore, to effectively leverage the dual structure of the problem, we also propose an optimal transport inspired alignment that serves to match cross-domain features. Specifically, we define a Gromov-Wasserstein distance \cite{memoli2011gromov} which measures the \emph{relational} distance between intra-domain distances while preserving the domain topological structures. Optimizing the Gromov-Wasserstein objective facilitates feature learning in each encoder network. It therefore promotes the proximity and similarity of the learnt embeddings despite inherent differences in domain topology.

Our contributions may be summarized as follows:
\begin{enumerate}
    \item We consider a new problem of both music-to-dance generation and its dual dance-to-music generation.
    \item A novel dual learning strategy is proposed, which incorporates a Gromov-Wasserstein distance to facilitate feature learning of each task, as well as promote coherence between input and output.
    \item Empirical experiments demonstrate the applicability of our approach in delivering realistic, diverse generations of dance choreographs and music compositions faithful to the input. Moreover, superior performance is observed when comparing to the state-of-the-arts in music-to-dance generation.
\end{enumerate}

\begin{figure*}[ht]
\includegraphics[width=\textwidth]{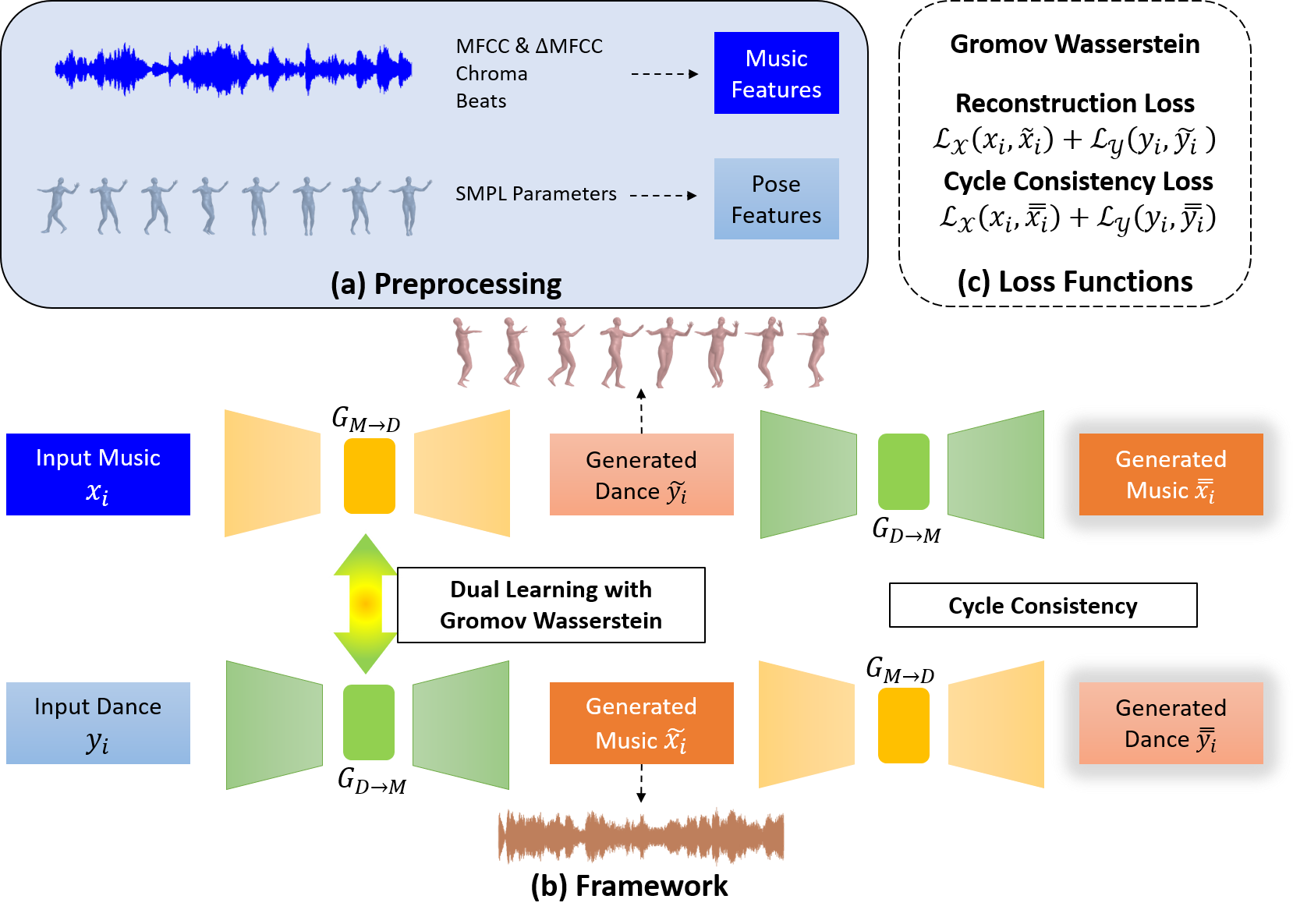}
\caption{A high-level overview of our pipeline. (a) For data preprocessing, we extract MFCC, chroma and beats features from the music waveform raw data, and represent the dance sequence as pose and translation parameters in the SMPL model \cite{loper2015smpl}. (b) The two generative networks $G_{M\to D}$ (music-to-dance choreography) and $G_{D\to M}$ (dance-to-music composition) in our framework comprise a sequence-to-sequence architecture where the encoder and decoder are both transformer networks. (c) We train our network through a Gromov-Wassserstein loss that facilitates dual learning in this cross-domain situation, on top of the reconstruction loss and cycle consistency loss.}
\label{fig:pipeline}
\end{figure*}

\section{Related Work}
\subsection{Dance generation}
Earlier works generate dance sequences from fixed templates following a similarity matching for the music source \cite{fan2011example,lee2013music}. Since the synthesized choreographs simply re-arrange the dance moves from the training data in rigid fashions, there is the drawback of unnatural transitions and a lack of creativity.

With the advent of deep learning methods, sequence-to-sequence models have been proposed for generating dance sequences from encoded music features. A pioneering work \cite{tang2018dance} uses a L2 loss for comparing the dance sequences, which suffers from a tendency of motion freezing. To alleviate this, \cite{huang2021dance} proposes curriculum learning coupled with a L1 loss whereas \cite{ye2020choreonet} employs a geodesic loss. To enable the generation of diverse dance sequences, \cite{huang2021dance,ye2020choreonet} introduce a random seed vector alongside the encoded music features, while an alternative approach \cite{lee2019dancing,ren2020self} utilises GANs. One key remark is that \cite{lee2019dancing,ren2020self,huang2021dance} focus on 2D motion. The 3D representation in our work allows important cues such as relative positions or invariance on bone lengths to be clearly put into perspective, thus appearing more more realistic, appealing and geometrically rich.

Another line of work \cite{li2020learning,li2021learn} employs a cross-modal architecture for generating dance sequences conditional on both music \emph{and} previous dance moves. In our work, the primary task of music-to-dance generation is only conditional on a single music stream and does not require any additional dance stream.

\subsection{Music generation}
There are two general approaches in computational music generation. The first focuses on symbolic representations \cite{briot2017deep,huang2018music,huang2020pop}. The second handles music as raw waveforms either in the audio \cite{dieleman2018challenge,child2019generating,dhariwal2020jukebox} or frequency domain \cite{vasquez2019melnet}. Ideally, the waveform representation enables a much richer generative landscape, such as allowing generation of human voices or nuances in musical performances which attach an additional layer of interpretation on top of the sheet music. However, this comes at an extremely high computational cost. To put this in context, the waveform representation for a 5 seconds music sequence sampled at 48kHz would incur a 240,000 length spectrogram sequence. Despite recent advancements such as sparse attentions \cite{child2019generating} or discretized representations \cite{dieleman2018challenge,dhariwal2020jukebox}, learning the multiple levels of musical structure and hierarchy which manifest at different scales remains extremely computationally expensive. In light of this, we resort to the lightweight symbolic representation for our music generation task.

\subsection{Dual Learning}
Dual learning \cite{xia2017dual} is a paradigm which jointly trains a primary task and its dual task. The symmetry and duality at the data or model level \cite{xia2018model} can be leveraged, ideally improving the performance of both tasks compared to training for each independently. Dual learning can be in a supervised setting such as in image-to-image translations for paired data \cite{isola2017image} or machines language translation \cite{he2016dual}. It may likewise be extended to the unsupervised setting \cite{yi2017dualgan,zhu2017unpaired} whereby the notion of a cycle consistency loss is introduced, compelling the primary task and dual task to learn inverse maps of each other. In our dual learning scheme, on top of the cycle consistency loss, we also introduce a novel Gromov-Wasserstein loss to facilitate feature space alignment.

\section{Our approach}
An overview of our pipeline is outlined in Fig. \ref{fig:pipeline}. In the following, we present the details of each component.

\subsection{Data Preprocessing}
\label{sec:preprocessing}
\textbf{Music} \quad The raw music input are sampled at 48kHz, thereby obtaining a waveform representation in the form of a time-frequency spectrogram. However, such high sampling rates may well lead to high computational costs and lots of redundancy. To this end, we employ the Librosa toolbox \cite{mcfee2015librosa} to perform feature engineering and extraction. These extracted features will be used in our networks instead of the raw waveform.

Following \cite{lee2019dancing}, we first extract Mel-frequency cepstral coefficients (MFCC) \cite{xu2004hmm} and MFCC delta from the spectrogram. However, these low level features (typically used for speech recognition) may be inadequate for conveying high level musical information. Thereafter, we perform a harmonic percussive source separation \cite{muller2015fundamentals} which decomposes the spectrogram into harmonic components and percussive components. The harmonic components correspond to the pitch and melody of the music from which we extract chroma features. The percussive components provide the rhythmic information from which we extract the beats and onsets. Specifically, for each frame, our music features is a 53 dimensional vector, comprising 20-dim MFCC, 20-dim MFCC delta, 12-dim chroma features, and 1-dim one-hot encoding for beats.

\textbf{Dance} \quad We work with two 3D dance sequence datasets from \cite{tang2018dance,li2021learn}. \cite{li2021learn} performed a 3D annotation of the AIST Dance database \cite{tsuchida2019aist} and each 3D pose is parameterized in the Skinned Multi-Person Linear Model \cite{loper2015smpl}. For consistency, we performed a fitting with SMPLify \cite{bogo2016keep} on the dataset in \cite{tang2018dance} to obtain SMPL pose parameters.

The SMPL model has $24$ skeletal joints and a pose is represented as $24\times3$ axis-angle parameters, which characterizes the 3D orientation or rotation of each joint. However the axis-angle (and likewise the quaternion) parameterization is not globally continuous over the 3D rotational group $SO(3)$ \cite{astey1987cobordism}. Furthermore, it is cumbersome to define a geometrically meaningful loss for axis-angle parameters and we adopt a Stiefel manifold representation \cite{zhou2019continuity}:
\begin{equation}\label{eqn:stiefelmanifold}
\begin{aligned}
\hat{R} &= 
\left(\begin{matrix} \vertbar & \vertbar \\ \mathbf{R}_1 & \mathbf{R}_2 \\ \vertbar & \vertbar \end{matrix}\right),
\end{aligned}
\end{equation}
which essentially amounts to discarding the last column for a rotation matrix. Such a representation is smooth over $SO(3)$, thus offering empirical advantages for backpropagation. Overall, the skeletal pose is parameterized as a $24\times6+3=147$-dim vector. 

\subsection{Problem Formulation}
We denote the music space as $\mathcal{X}$ and the dance space as $\mathcal{Y}$. We sample fixed duration ($T$ frames) music and dance sequence pairs $(x_i,y_i)_{i=1}^N$ \emph{i.i.d} from $\mathcal{X}\times\mathcal{Y}$. Formally, the problem statement can be formulated as follows.
\begin{itemize}
\item The primary task is to learn a mapping $G_{M\to D}:\mathcal{X}\to\mathcal{Y}$ such that $\sum_{i=1}^N\mathcal{L}_{\mathcal{Y}}(G_{M\to D}(x_i),y_i)$ is minimized. Here $\mathcal{L}_{\mathcal{Y}}$ denotes a metric in dance space $\mathcal{Y}$.
\item The dual task is to learn a mapping $G_{D\to M}:\mathcal{Y}\to\mathcal{X}$ such that $\sum_{i=1}^N\mathcal{L}_{\mathcal{X}}(G_{D\to M}(y_i),x_i)$ is minimized. Here $\mathcal{L}_{\mathcal{X}}$ denotes a metric in music space $\mathcal{X}$.
\end{itemize}
In the ideal case, $G_{M\to D}$ and $G_{D\to M}$ would be inverse of each other, \emph{i.e.} $G_{D\to M}(G_{M\to D}(x_i))=x_i$ and $G_{M\to D}(G_{D\to M}(y_i))=y_i$. The discrepancy from this ideal case may be leveraged in the form of a cycle discrepancy loss \cite{zhu2017unpaired} in our dual learning context.

\begin{figure}[ht]
\includegraphics[width=\linewidth]{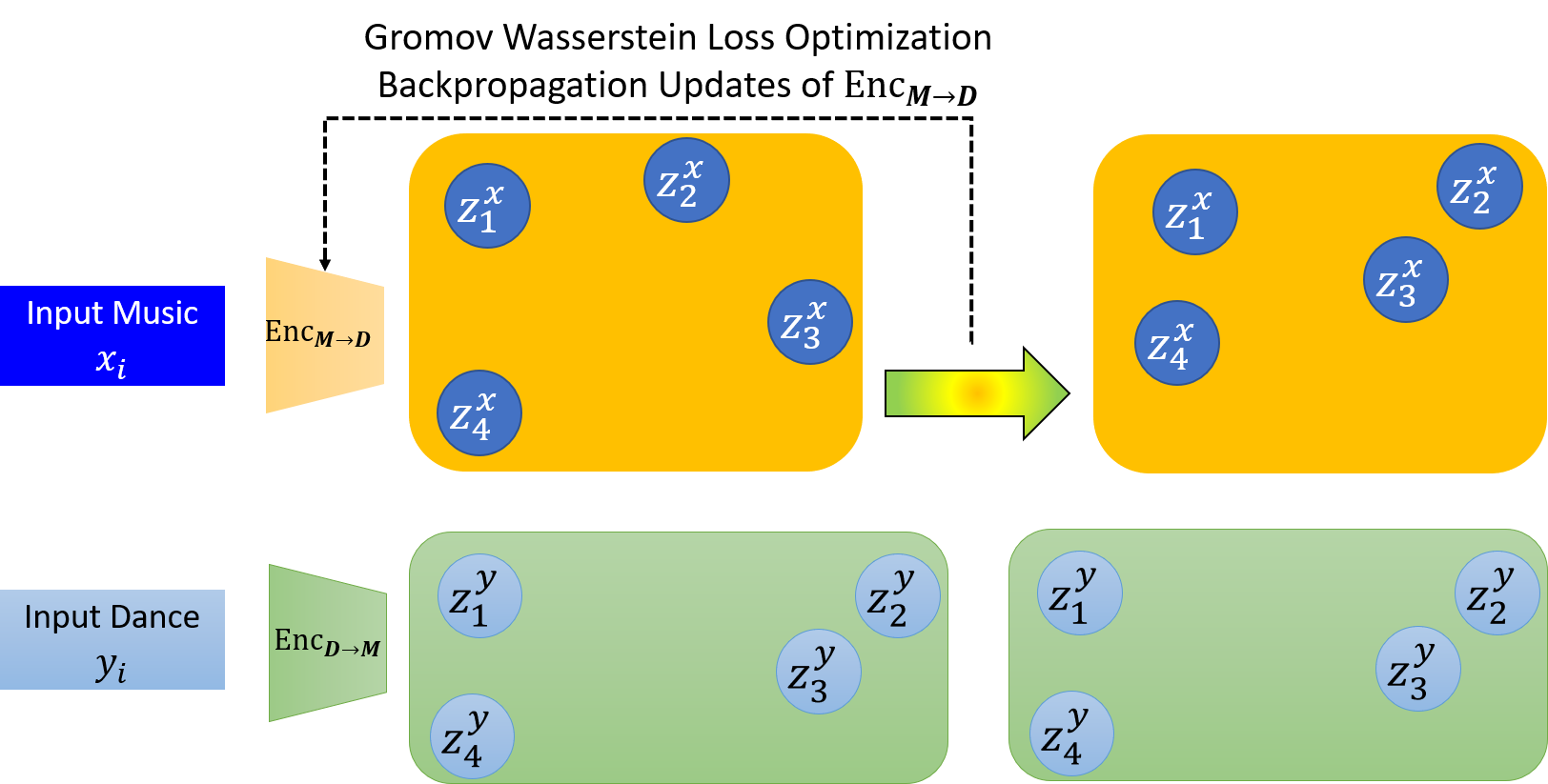}
\caption{In optimizing for the Gromov-Wasserstein loss, the $\text{Enc}_{M\to D}$ network parameters are updated, re-positioning the music embedding vectors $\{z^x_i\}_{i=1}^4$ such that the discrepancy between intra-space distances is minimized.}
\label{fig:gw}
\end{figure}

\subsection{Overall Framework}
\label{sec:framework}
\textbf{Transformer and attention} \quad Both the $G_{M\to D}$ and $G_{D\to M}$ consist of transformer networks that serve as encoders and decoders as illustrated in Fig. \ref{fig:pipeline}. We adopt a full attention mechanism \cite{devlin2018bert} throughout for all transformers. This allows full access to the entire context, without masking of the future contexts. Explicitly, for a given sequence input $\mathbf{S}$, an attention layer learns the context $\mathbf{Z}$:
\begin{equation}\label{eqn:attention}
\begin{aligned}
\mathbf{Z} = \text{Softmax}\left(\frac{\langle\mathbf{S}\mathbf{W}_q,\mathbf{S}\mathbf{W}_k\rangle}{\sqrt{D}}\right)\mathbf{S}\mathbf{W}_v,
\end{aligned}
\end{equation}
where $D$ is the number of channels in the attention layer and $\mathbf{W}_q,\mathbf{W}_k,\mathbf{W}_v$ denote the query, key and value weights respectively.

Intuitively, the full attention mechanism aligns with how humans go about composing music for dance or choreographing dance from music. We make a global consideration of the structure and themes in our design. In allowing access to all contextual information, we expect that the structural constructs and hierarchical abstractions can be better modeled in the transformer networks.

\textbf{Gromov-Wasserstein Loss to Facilitate Dual Learning in Encoders} \quad Each encoder network learns an embedding map from its respective input space into a latent space. Explicitly, we have:
\begin{equation}\label{eqn:embedding}
\begin{aligned}
\text{Enc}_{M\to D}: \mathcal{X} &\to \mathcal{Z}_\mathcal{X}\\
x&\mapsto z^x,\\
\text{Enc}_{D\to M}: \mathcal{Y} &\to \mathcal{Z}_\mathcal{Y}\\
y&\mapsto z^y.
\end{aligned}
\end{equation}
The music space $\mathcal{X}$ and dance space $\mathcal{Y}$ have different topological structures. As such, it would be mathematically impossible to embed them into the same space since embeddings are topology-preserving maps. This means that we cannot directly gauge the similarity between a piece of music and a dance sequence by defining a metric on their embedding vectors.

On the other hand, embeddings do introduce metric distortions \cite{bourgain1985lipschitz}. This motivates us to propose a Gromov-Wasserstein loss \cite{memoli2011gromov,peyre2016gromov,alvarez2018gromov} which measures the discrepancy between the metrics on the two embedding spaces instead of directly comparing between cross-domain samples. By defining an \emph{inter-space} distance between the respective \emph{intra-space} distances for $\mathcal{Z}_\mathcal{X}$ and $\mathcal{Z}_\mathcal{Y}$, the Gromov-Wasserstein loss gives a well-defined notion of distance between music and dance pairs. Heuristically, we present in Fig. \ref{fig:gw} how optimizing this loss amounts to aligning the feature embeddings.

Give two point sets of $m$ embedded vectors $\{z^x_i\}_{i=1}^m,\{z^y_i\}_{i=1}^m$, we may view them as two discrete empirical distributions $\mu,\nu$ as:
\begin{equation}
\begin{aligned}
\mu = \sum_{i=1}^m\frac{1}{m}\delta_{z^x_i}, \;\; \nu = \sum_{i=1}^m\frac{1}{m}\delta_{z^y_i},
\end{aligned}
\end{equation}
where $\delta$ denotes the Dirac delta distribution. Formally, the Gromov-Wasserstein distance for our task is given by
\begin{equation}\label{eqn:gw}
\begin{aligned}
GW(\mu,\nu)=\min_{\pi\in\Pi}\sum_{i,j,k,l} \left\lvert  \lVert z^x_i-z^x_k \rVert_1 -  \lVert z^y_j-z^y_l  \rVert_1 \right\rvert^2 \pi_{ij} \pi_{kl}.
\end{aligned}
\end{equation}
Here $\Pi$ defines the set of all joint distributions with marginals $\mu$ and $\nu$. The goal is to find the optimal transport matrix $\pi$ that minimizes the squared distance between intra-space L1 costs.

Following \cite{peyre2016gromov}, we introduce an entropic regularization term which allows much more efficient solving of Eq.~\eqref{eqn:gw}. The entropy regularized Gromov-Wasserstein distance can then be solved via the Sinkhorn algorithm and projected gradient descent \cite{peyre2016gromov}. We outline the steps in Algorithm~\ref{algo:GW}.
\begin{algorithm}
\caption{GW Distance for 2 batches of $m$ samples}
\label{algo:GW}
\begin{algorithmic}
\renewcommand{\algorithmicrequire}{\textbf{Input:}}
\Require music embeddings $\mathbf{Z}^x=\{z^x_i\}_{i=1}^m,\mathbf{Z}'^x=\{z'^x_i\}_{i=1}^m$
\Require dance embeddings $\mathbf{Z}^y=\{z^y_i\}_{i=1}^m,\mathbf{Z}'^y=\{z'^y_i\}_{i=1}^m$
\renewcommand{\algorithmicrequire}{\textbf{Hyperparameters:}}
\Require regularization $\varepsilon > 0$, projection iterations $M$, Sinkhorn iterations $L$
\renewcommand{\algorithmicrequire}{\textbf{Initialize:}}
\Require $\pi^{(0)}_{ij} = \frac{1}{n}, \; \forall{i,j}$
\State Cost Matrix for music embeddings $C_{ij}=\left\lVert z^x_i-z^x_j \right\rVert_1$
\State Cost Matrix for dance embeddings $D_{ij} = \left\lVert z^y_i-z^y_j \right\rVert_1$
\For{$l = 1:M$}
\State $E = \frac{1}{m}D^2 \mathds{1}_m \mathds{1}_m^\intercal + \frac{1}{m} \mathds{1}_m \mathds{1}_m^\intercal C^2 - 2D \pi^{(l-1)} C^\intercal$
\State $K = \exp(-E/\varepsilon)$
\State $\mathbf{b}^{(0)}  = \mathds{1}_m$
\For{$\ell = 1:L$}
\State $\mathbf{a}^{(\ell)} = \mathds{1}_m \oslash K\mathbf{b}^{(\ell-1)}$, \; $\mathbf{b}^{(\ell)} = \mathds{1}_m \oslash K^\intercal\mathbf{a}^{(\ell)}$
\State where $\oslash$ denotes component-wise division
\EndFor
\State $\pi^{(l)}=\text{diag}(\mathbf{a}^{(L)})K\text{diag}(\mathbf{b}^{(L)})$
\EndFor
\Ensure \small$GW_{\varepsilon}(\mathbf{Z}^x,\mathbf{Z}'^x,\mathbf{Z}^y,\mathbf{Z}'^y) = \displaystyle \sum_{i,j,k,l} \left\lVert D_{ik}-C_{jl} \right\rVert^2\pi_{ij}^{(M)}\pi_{kl}^{(M)}$
\end{algorithmic}
\end{algorithm}

Optimizing the Gromov-Wasserstein distance through Algorithm \ref{algo:GW} updates the music encoder network parameters $\text{Enc}_{M\to D}$ through backpropagation. This matching of the music embeddings with the dance embeddings facilitates learning the duality of our two tasks.

\textbf{Decoding and Generation} \quad During training phase, the start token for the decoder network is taken from the training data. Specifically, for the dance generation task, the start token is the initial dance pose. During inference phase, we sample a random seed vector for the initial pose instead. For the music generation task, during inference, we randomly sample a chord root from \{C,C\#,D,D\#,E,F,F\#,G,G\#,A,A\#,B\} and a chord quality among \{major, minor\}. These random seeds during the inference phase introduce diversity in the generated dance and music.

\subsection{Loss Functions}
\label{sec:loss}
As illustrated in Fig. \ref{fig:pipeline}, three loss functions are defined. The first is our Gromov-Wasserstein loss, which has been discussed in detail in the previous subsection. Essentially it serves as an auxiliary regularization loss that promotes a better learning of the correspondence between the music and dance embeddings. This facilitates extracting the commonalities and shared structural similarities between the two media, which would enhance the consistency between inputs and generated outputs in our tasks.

The reconstruction loss has the general form:
\begin{equation}\label{eqn:reconstruction}
\begin{aligned}
\mathcal{L}^\text{reconstruction}_{\text{dance}}&=&\sum_i\mathcal{L}_{\mathcal{Y}}(y_i,G_{M\to D}(x_i)),\\
\mathcal{L}^\text{reconstruction}_{\text{music}}&=&\sum_i\mathcal{L}_{\mathcal{X}}(x_i,G_{D\to M}(y_i)).
\end{aligned}
\end{equation}
$\mathcal{L}_{\mathcal{Y}}$ denotes the metric over the dance space $\mathcal{Y}$. Specifically, we have $\mathcal{Y}=(\mathbb{R}^3\times \underbrace{SO(3)\times\cdots\times SO(3)}_{24\text{ times}})^T$ where $T$ is the total number of frames in our model. A point $y\in\mathcal{Y}$ may then be written as $y=\left[y^{\text{trans}}_t,y^{\text{rot}_1}_t,\cdots,y^{\text{rot}_{24}}_t\right]_{t=1}^T$. We define $\mathcal{L}_{\mathcal{Y}}:\mathcal{Y}\times\mathcal{Y}\to\mathbb{R}_+$ as follows:
\begin{flalign}\label{eqn:dancemetric}
&\small{\mathcal{L}_{\mathcal{Y}}(y,\tilde{y})=\sum_{t=1}^T\underbrace{\lVert y^{\text{trans}}_t-\tilde{y}^{\text{trans}}_t\rVert_1}_{\text{L1 loss for translation}}+\sum_{t=1}^T\sum_{j=1}^{24}\text{geodesic}(y^{\text{rot}_j}_t,\tilde{y}^{\text{rot}_j}_t)^2}.
\end{flalign}
Here, $\text{geodesic}:SO(3)\times SO(3)\to\mathbb{R}_+$ defines the shortest distance between two 3D rotations. Recall that $y^{\text{rot}_j}_t$ constitutes the first two columns of its associated rotation matrix $R^j_t$ as defined in Eq.~\eqref{eqn:stiefelmanifold}, we can recover its third column through the cross product. The geodesic distance between rotation matrices $R,\tilde{R}$ is given by:
\begin{flalign}\label{eqn:geodesic}
\text{geodesic}(R,\tilde{R}) = \left\lvert \arccos \left[\frac{\text{Tr}(R\tilde{R}^\intercal)-1}{2}\right]\right\rvert.
\end{flalign}

For the music space $\mathcal{X}$, we compare only the chroma and beats. The metric $\mathcal{L}_{\mathcal{X}}:\mathcal{X}\times\mathcal{X}\to\mathbb{R}_+$ is defined as:
\begin{flalign}\label{eqn:musicmetric}
&\mathcal{L}_{\mathcal{X}}(x,\tilde{x})=\sum_{t=1}^T\underbrace{\lVert x^{\text{chroma}}_t-\tilde{x}^{\text{chroma}}_t\rVert_1}_{\text{L1 loss for chroma}}+\underbrace{\lVert x^{\text{beats}}_t-\tilde{x}^{\text{beats}}_t\rVert_1}_{\text{L1 loss for beats}}.
\end{flalign}
The cycle consistency loss is another regularization term that we incorporate to facilitate dual learning. Whereas the Gromov-Wasserstein loss is applied at an initial stage to update the music embedding network parameters, the cycle consistency loss is applied at the posterior stage to gauge the generated rendition against the input. With the same metrics defined in Eqs.~\eqref{eqn:dancemetric} and \eqref{eqn:musicmetric}, the cycle consistency loss measures the discrepancy of the two dual networks $G_{M\to D}$ and $G_{D\to M}$ from being inverse to each other:
\begin{equation}\label{eqn:cycleconsistency}
\begin{aligned}
\mathcal{L}^\text{cycle}_{\text{dance}}&=&\sum_i\mathcal{L}_{\mathcal{Y}}(y_i,G_{M\to D}(G_{D\to M}(y_i))),\\
\mathcal{L}^\text{cycle}_{\text{music}}&=&\sum_i\mathcal{L}_{\mathcal{X}}(x_i,G_{D\to M}(G_{M\to D}(x_i))).
\end{aligned}
\end{equation}
Our transformer networks $G_{M\to D}$ and $G_{D\to M}$ are trained concurrently and updated according to the following prescriptions: $$\mathcal{L}^\text{GW}+\mathcal{L}^\text{reconstruction}_{\text{dance}}+\mathcal{L}^\text{cycle}_{\text{dance}}\xrightarrow[\text{}]{\text{backpropagation}}G_{M\to D},$$
$$\mathcal{L}^\text{reconstruction}_{\text{music}}+\mathcal{L}^\text{cycle}_{\text{music}}\xrightarrow[\text{}]{\text{backpropagation}}G_{D\to M}.$$

\begin{figure*}[t]
\centering
\resizebox{\textwidth}{!}{
\begin{tabular}{|l||c||c|}
\hline
Method & Click $\downarrow$ & Cha Cha dance sequences displayed at 0.2 second intervals
\\ \hline
Ground Truth &
\animategraphics[height=1.2cm]{5}{Figures/GT/}{0}{49} & \includegraphics[height=1.2cm]{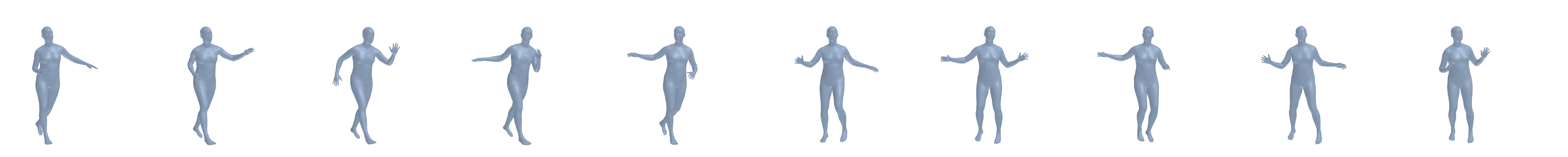}
\\ \hline
Tang et al. (2018) \cite{tang2018dance} &
\animategraphics[height=1.2cm]{5}{Figures/Tang(2018)/}{0}{49} & \includegraphics[height=1.2cm]{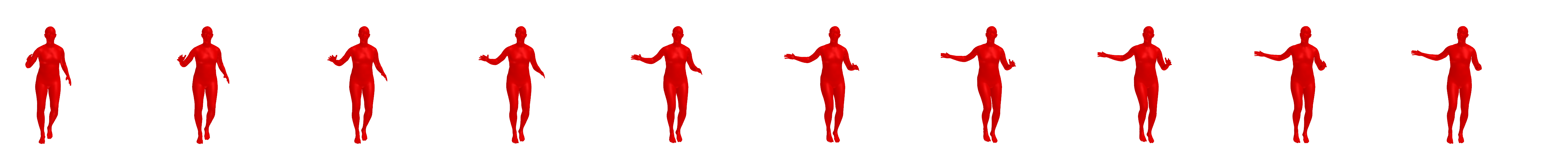}
\\ \hline
Ren et al. (2020) \cite{ren2020self} &
\animategraphics[height=1.2cm]{5}{Figures/Ren(2020)/}{0}{49} & \includegraphics[height=1.2cm]{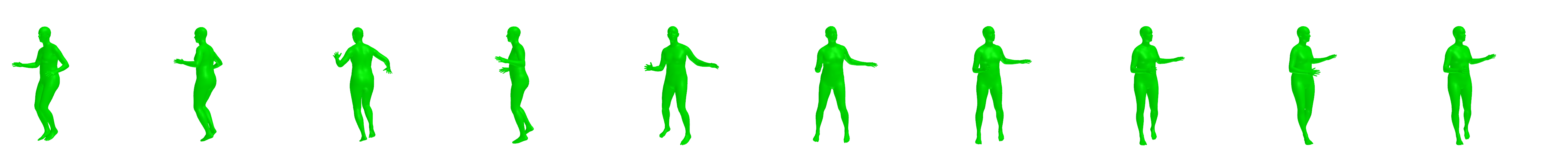}
\\ \hline
Huang et al. (2021) \cite{huang2021dance} &
\animategraphics[height=1.2cm]{5}{Figures/Huang(2021)/}{0}{49} & \includegraphics[height=1.2cm]{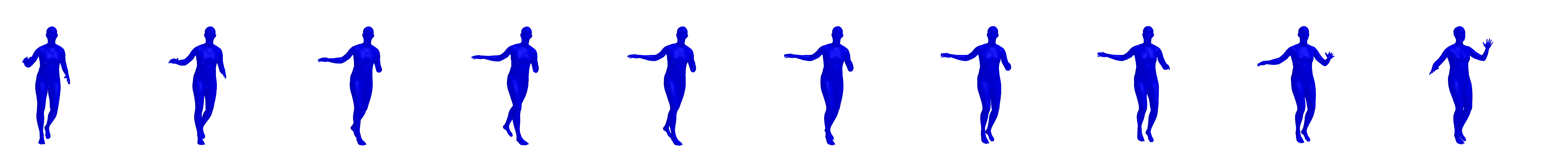}
\\ \hline
Ours &
\animategraphics[height=1.2cm]{5}{Figures/Ours/}{0}{49} & \includegraphics[height=1.2cm]{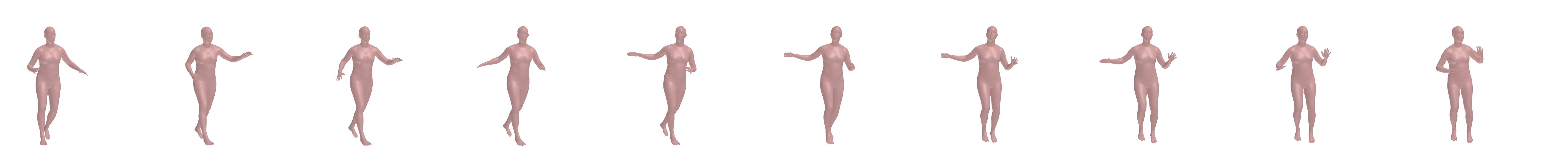}
\\ \hline
\end{tabular}
}
\caption{Visual comparison of sample generated Cha Cha dance sequences. Best viewed via Adobe Acrobat Reader. Click the figures under `Click $\downarrow$' to show dance animations (4 seconds at 5 FPS).}
\label{fig:realism}
\end{figure*}
\section{Experiments}
\subsection{Dataset Description}
As mentioned in subsection \ref{sec:preprocessing}, we employ two public datasets from \cite{tang2018dance} and \cite{tsuchida2019aist,li2021learn}, which consists of 3D dance sequences with accompanying music. Since these two datasets do not overlap in their dance genres, we combine them into a single entity. We first re-parameterized the skeletal pose representation in \cite{tang2018dance} in the SMPL model \cite{loper2015smpl} pose parameters (3D positional representation of 17 keypoints $\to$ rotational pose representation of 24 keypoints) through an inverse kinematics fitting. For consistency in frame rate, we also perform a downsampling (via spherical interpolation) of the frames per second (FPS) in \cite{li2021learn} from 60 to 25. The statistics for the aggregated dataset is summarized in the following table.
\begin{table}[h]
\resizebox{1\linewidth}{!}{
\begin{tabular}{|l|c|c|c|}
\hline
Dance Genre & \# of Sequences & Frames & Remarks \\\hline
Rumba & 10 & 20950 & \multirow{4}{*}{\parbox{3.2cm}{From \cite{tang2018dance}. Dance sequences are typically much longer at around 150 seconds.}} \\\cline{1-3}
Cha Cha & 8 & 20425 & \\\cline{1-3}
Tango & 9 & 49165 & \\\cline{1-3}
Waltz & 34 & 43298 & \\\hline
Break Dance & 141 & 46526 & \multirow{10}{*}{\parbox{3.2cm}{From \cite{li2021learn}. Dance sequences generally range from 8 to 12 seconds.}} \\ \cline{1-3}
House & 141 & 40050 & \\\cline{1-3}
Ballet Jazz & 141 & 47727 & \\\cline{1-3}
Street Jazz & 141 & 47920 & \\\cline{1-3}
Krump & 141 & 47534 & \\\cline{1-3}
LA Style Hip Hop & 141 & 48323 & \\\cline{1-3}
Lock & 141 & 47388 & \\\cline{1-3}
Middle Hip Hop & 141 & 48276 & \\\cline{1-3}
Pop & 140 & 46749 & \\\cline{1-3}
Waack & 140 & 47355 & \\\hline
\end{tabular}}
\caption{Summary of dataset, aggregated from \cite{tang2018dance} and \cite{tsuchida2019aist,li2021learn}. For our tasks, the frame rate is set to 25 FPS. During training, the input and output sequences are fixed at 75 frames (or 3 seconds).}
\vspace{-2em}
\label{tab:dataset}
\end{table}

\subsection{Implementation Details}
Our framework is implemented with PyTorch. We set the input and output sequence lengths to $T=75$ frames (equivalent to 3 seconds). Both $G_{M\to D}$ and $G_{D\to M}$ have 6 transformer layers and 8 attention heads. The hidden units dimensions for the transformer layers is set to 512 for $G_{M\to D}$ and 256 for $G_{D\to M}$. For our Gromov-Wasserstein loss, the hyperparameters for Algorithm \ref{algo:GW} are as follows: the entropic regularization parameter is set to $\varepsilon=0.2$ while the number of Sinkhorn iterations and projection iterations are respectively set to $L=30$ and $M=20$. During training, we use the Adam optimizer with a batch size of 16 and an initial learning rate of 1e-4 (decays to \{1e-5, 5e-6\} after \{20k, 40k\} iterations).

During inference, the decoder start token may either be sampled randomly from a pool (for diverse generation setting) or provided as an auxiliary input. For inference over 75 frames, we simply feed the last generated frame as the decoder start token. 

\subsection{Dance Generation}
For generation of music-to-dance choreography, we compare our work against \cite{tang2018dance,ren2020self,huang2021dance}. For \cite{tang2018dance}, we re-implement it since the source code is not available. We also adapt \cite{ren2020self,huang2021dance} (both implemented for 2D dance sequences) for 3D dance generation. We employ three quantitative metrics, namely the Fr\'{e}chet Distance, Diversity and Beats Alignment. The details of these metrics are discussed below and the quantitative results are reported in Tab. \ref{tab:dance_quantitative}. 

Furthermore, we engage a qualitative user study to rate the realism and the genre consistency of the generated dances. The results are tabulated in Tab. \ref{tab:dance_user}.

\begin{figure*}[t]
\centering
\resizebox{\textwidth}{!}{
\begin{tabular}{|c||c||c|}
\hline
Genre & Click $\downarrow$ & Dance sequences displayed at 0.2 second intervals
\\ \hline
\multirow{3}{3cm}{Cha Cha \includegraphics[width=3cm]{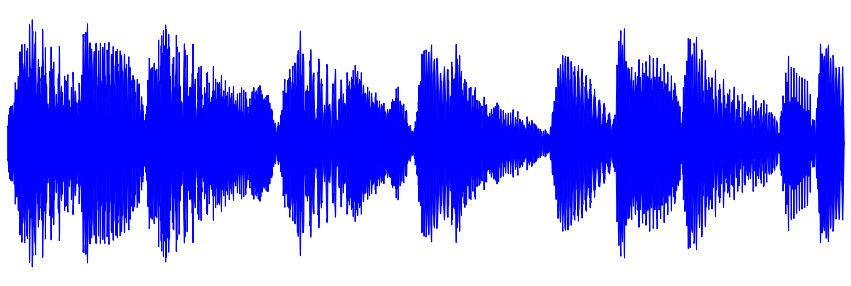}} &
\animategraphics[height=1.2cm]{5}{Figures/Diversity_ChaCha1/}{0}{49} & \includegraphics[height=1.2cm]{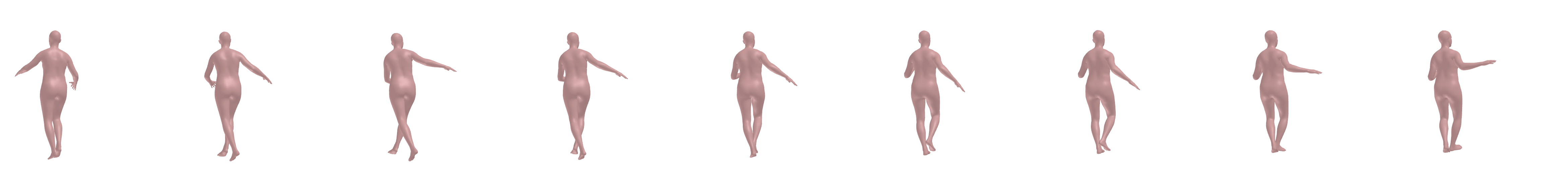}
\\ \cline{2-3}
 &
\animategraphics[height=1.2cm]{5}{Figures/Diversity_ChaCha2/}{0}{49} & \includegraphics[height=1.2cm]{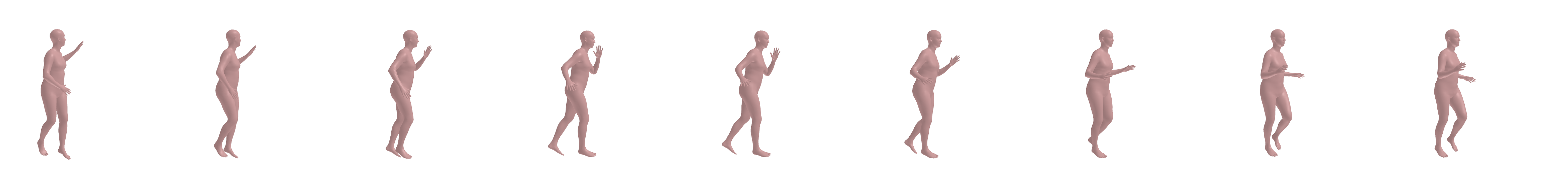}
\\ \cline{2-3}
 &
\animategraphics[height=1.2cm]{5}{Figures/Diversity_ChaCha3/}{0}{49} & \includegraphics[height=1.2cm]{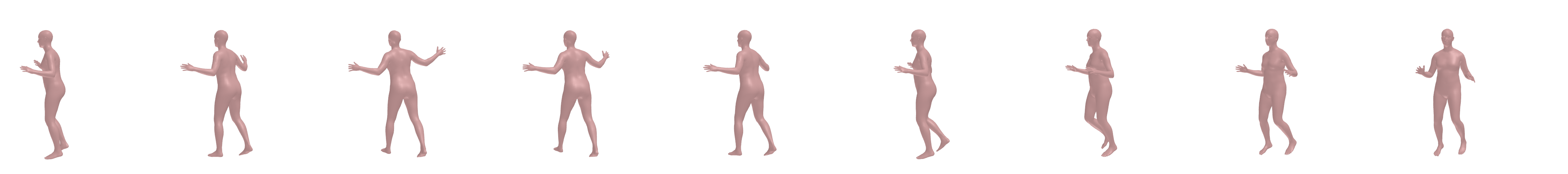}
\\ \hline
\multirow{3}{3cm}{Break Dance \includegraphics[width=3cm]{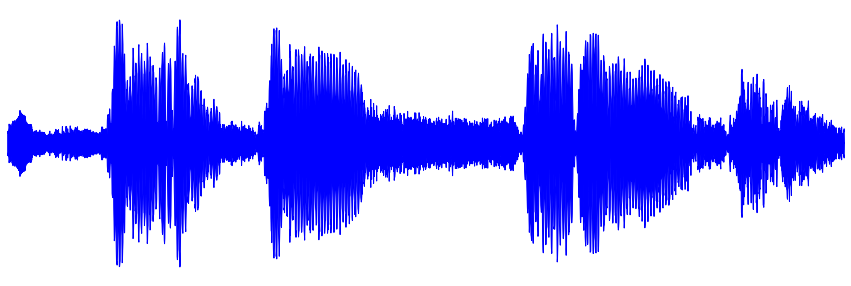}} &
\animategraphics[height=1.2cm]{5}{Figures/Diversity_Break1/}{0}{49} & \includegraphics[height=1.2cm]{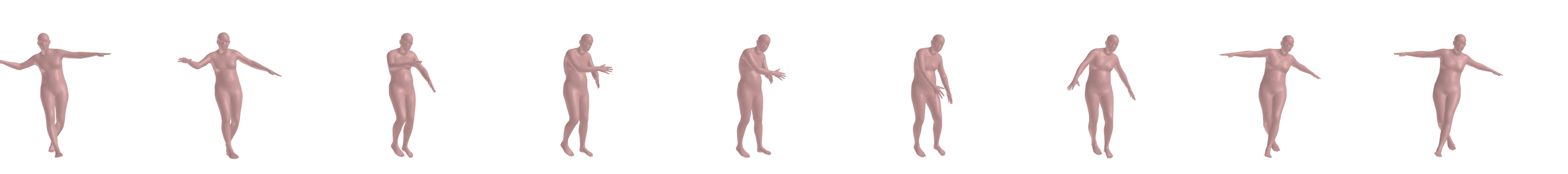}
\\ \cline{2-3}
 &
\animategraphics[height=1.2cm]{5}{Figures/Diversity_Break2/}{0}{49} & \includegraphics[height=1.2cm]{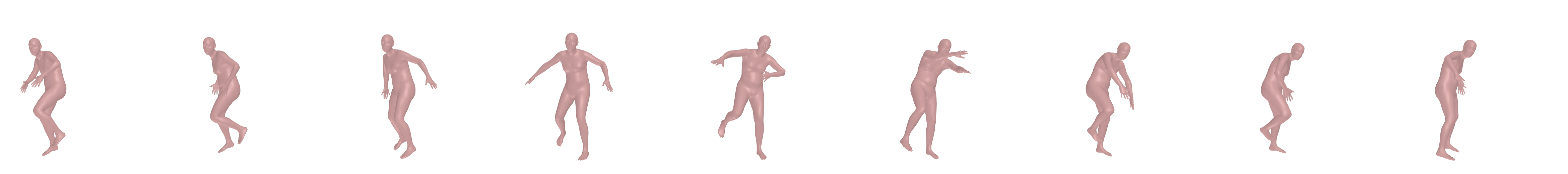}
\\ \cline{2-3}
 &
\animategraphics[height=1.2cm]{5}{Figures/Diversity_Break3/}{0}{49} & \includegraphics[height=1.2cm]{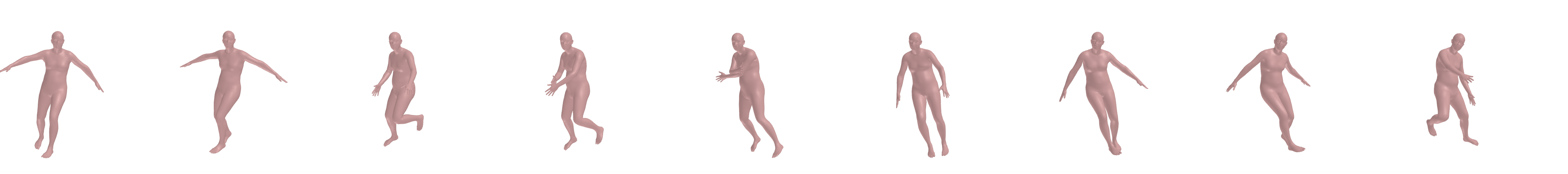}
\\ \hline
\end{tabular}
}
\caption{Our framework can generate diverse dance sequences given the same input music but a different starting pose. Best viewed via Adobe Acrobat Reader where dance animations will be played upon clicking the figures under `Click $\downarrow$'.}
\label{fig:diversity}
\end{figure*}

We showcase sample dance animations in Fig. \ref{fig:realism} and \ref{fig:diversity}. These sequences spanning 10 seconds are animated at 5 FPS (requires Adobe Acrobat Reader). More sample sequences are available in the supplementary video.

\begin{table}[h]
\resizebox{1\linewidth}{!}{
\begin{tabular}{|l|c|c|c|}
\hline
Method & Fr\'{e}chet Distance & Diversity & Beats Alignment (\%) \\ \hline
Ground Truth & - & - & 68.7 \\ \hline
Tang et al. (2018) \cite{tang2018dance} & 986.4 & 10.3 & 31.2 \\ \hline
Ren et al. (2020) \cite{ren2020self} & 1526.3 & 48.2 & 49.5 \\ \hline
Huang et al. (2021) \cite{huang2021dance} & 384.2 & 37.2 & 62.3 \\ \hline
Ours & \textbf{140.5} & \textbf{49.8} & \textbf{64.5} \\ \hline
\end{tabular}}
\caption{Quantitative results for dance generation. We adopt three metrics: 1) \emph{Fr\'{e}chet distance} measures the difference from the ground truth, 2) \emph{Diversity} measures the variation in dance moves, and 3) \emph{Beats Alignment} evaluates the percentage matching of kinematic beats and music beats. Best performance highlighted in bold.}
\vspace{-2em}
\label{tab:dance_quantitative}
\end{table}

\textbf{Fr\'{e}chet Distance} \quad
We define the Fr\'{e}chet distance as the average 3D joint distance of the generated dance sequence from the ground truth. In our experiments, we evaluate the Fr\'{e}chet distance for 280 sequences (20 per genre, each spanning 9 seconds) in our pre-allocated test set. Our framework achieves significantly better performance than compared methods. This superior matching of the generated dances with the ground truth suggests that the music-to-dance correspondence is better modeled in our approach.

Sample generated Cha Cha sequences are shown in Fig. \ref{fig:realism} for the same initial pose and input music. We observe that our network generates sequences more consistent with the ground truth. \cite{tang2018dance} has a inclination to gravitate towards motionless states, showing very limited range of motions. On the other hand, while not suffering from lack of motions, \cite{ren2020self} tends to deviate from the ground truth.

\begin{table}[h]
\resizebox{1\linewidth}{!}{
\begin{tabular}{|l|c|c|}
\hline
Method & Realism (Ranking) & Genre Consistency (Ranking) \\ \hline
Tang et al. (2018) \cite{tang2018dance} & 3.9 & 3.80 \\ \hline
Ren et al. (2020) \cite{ren2020self} & 2.98 & 2.58 \\ \hline
Huang et al. (2021) \cite{huang2021dance} & 2.05 & 2.35 \\ \hline
Ours & \textbf{1.07} & \textbf{1.27} \\ \hline
\end{tabular}}
\caption{User study for dance generation. We report the average ranking of each method for motion realism and dance genre consistency.}
\vspace{-3em}
\label{tab:dance_user}
\end{table}

\begin{figure*}[t]
\centering
\includegraphics[width=0.95\textwidth]{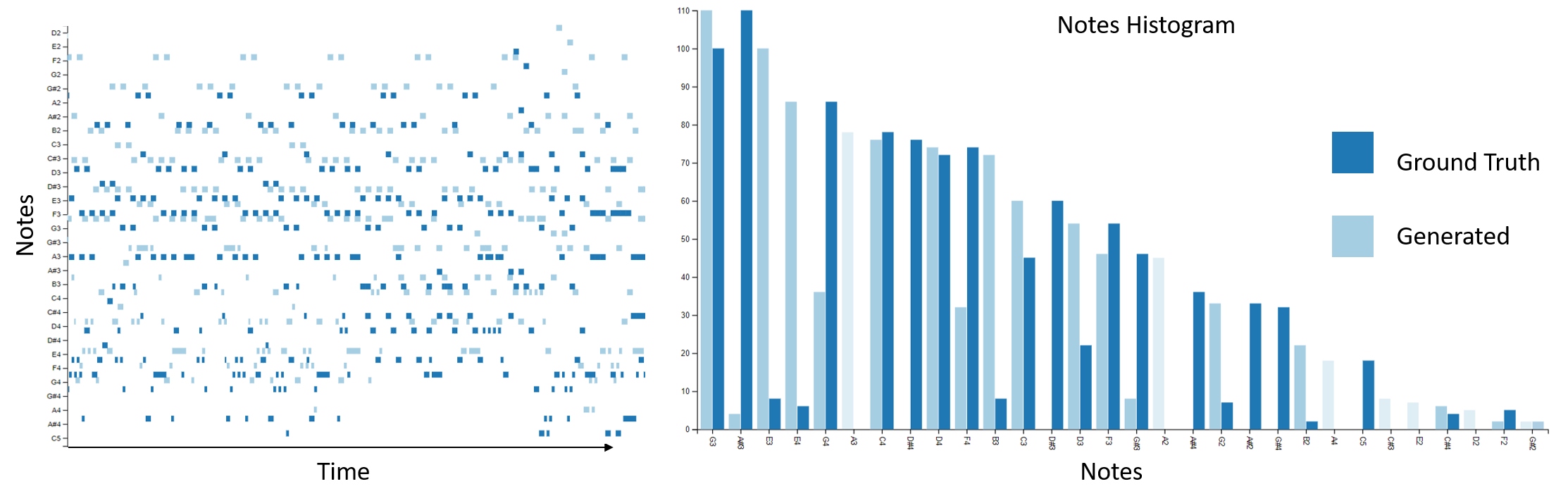}
\caption{Visualization of sample generated music (light blue) and original music (dark blue). Left: Visualization of music notes played out in time. Right: Histogram of music notes.}
\label{fig:music}
\end{figure*}

\textbf{Diversity} \quad
Recall that we incorporate randomness into our decoder through a randomly seeded start token. This allows our framework to generate diverse dance moves from the same input music but a different initial pose. We define our diversity metric as the variation in 3D joints position, evaluated over 5 generated dance sequences conditioned on the same music and different initial poses. This is averaged over 20 independent trials for each genre. In Fig. \ref{fig:diversity}, We showcase the dance choreographs obtained by our model given the same input music. The diverse generation is indicative that our model is adept at learning the abstract movements in music and translating them to kinematic movements in dance.

\textbf{Beats Alignment} \quad
Beats alignment measures the consistency of rhythmic articulation. For simplicity, we follow \cite{li2021learn} in defining dance kinematics beats as local minima in average joints 3D speed. We then define a matching beat if the occurrence of the kinematics beat occurs within 5 frames or 0.2 seconds of a music beat. The beats alignment ratio can then be defined as the ratio of matching beats to the total dance beats. Our framework again achieves compelling performance for beats matching, demonstrating its effectiveness in learning the rhythmic articulation of the input music.

\textbf{User Study} \quad
We engage 5 users (amateur dancers) to assess the generated dances in a single blind study. For each dance genre, we generate 3 dance sequences of 12 seconds duration via different methods. Each user then rank the generated dances according to two criteria: 1) the naturalness or realism of the dance poses; 2) the consistency with the dance genre. As shown in Tab. \ref{tab:dance_user}, the dance sequences generated from our framework is overwhelmingly ranked most preferred for both realism and genre consistency.

\subsection{Music Generation}
To our knowledge, we are the first to investigate the task of composing music for given dance sequences. We propose to quantify the performance of our model via a notes accuracy metric. We first transpose both compared music pieces to either the C major or A minor scale (depending on its original chord quality), before computing the average notes accuracy. Our model achieves an accuracy of 72\%, as tabulated in Tab. \ref{tab:ablation}.

However, this metric is not ideal since music composition is rather subjective task and is best assessed via human listening tests \cite{huang2020pop}. To this end, we invite the reader to look over and listen to the sample music compositions in our supplementary file. A sample generated music piece, is shown in tandem with the ground truth in Fig. \ref{fig:music}. The visualization is done with JuxtaMidi \cite{JuxtaMidi}.

\subsection{Ablation Studies}
\begin{table}[h]
\resizebox{1\linewidth}{!}{
\begin{tabular}{|c|c|c|c|c|c|}
\hline
\multirow{2}{*}{GW} & \multirow{2}{*}{CC} & \multicolumn{3}{c|}{Dance} & Music \\ \cline{3-6}
& & Fr\'{e}chet Distance & Diversity & Beats Alignment (\%) & Notes Accuracy \\ \hline
 & \checkmark & 335.4 & 46.3 & 58.5 & 0.51 \\ \hline
\checkmark & & 196.2 & \textbf{50.1} & 60.3 & 0.59 \\ \hline
 & & 376.5 & 46.2 & 56.4 & 0.37 \\ \hline
\checkmark & \checkmark & \textbf{140.5} & 49.8 & \textbf{64.5} & \textbf{0.72} \\ \hline
\end{tabular}}
\caption{Ablation experiments performed on two components of our framework. \emph{GW} refers to the Gromov-Wasserstein loss and \emph{CC} refers to the cycle consistency loss.}
\vspace{-3em}
\label{tab:ablation}
\end{table}

We perform ablation studies focused on investigating the effectiveness of our dual learning scheme. The first key component in our framework is the Gromov-Wasserstein loss, applied at the early stage of a training iteration to align the music embeddings with dance embeddings. The second component is the cycle consistency loss, optimized at the final stage of a training iteration as an additional regularization to enforce overall consistency.

The ablation experiments are reported in Tab. \ref{tab:ablation}. We observe significant performance drops for the Fr\'{e}chet distance, beats alignment ratio and music notes accuracy upon removal of the Gromov-Wasserstein objective. This provides strong justification of its potency for handling the duality of the tasks. The cycle consistency loss, is also effective, albeit to a lesser extent. Overall, compared to independently training separate task-specific networks, the conjoined learning in our framework delivers significant improvements through a balanced set-up of well-tailored components.

\section{Conclusion}
In this work, we propose a novel problem of simultaneously learning music-to-dance choreography and dance-to-music composition. The crucial ingredient is how to effectively leverage the duality of the tasks and integrate the information from both domains. To overcome the challenge of our cross-domain setting, we design a Gromov-Wasserstein objective for aligning the music embeddings vis-\`{a}-vis the dance embeddings, coupled to a cycle consistency loss. These auxiliary losses and our dual learning scheme prove capable in boosting the performance of individual tasks. Our framework delivers realistic and genre-consistent dance generations, as well as viable music compositions. For future work, we seek to extend our framework to raw waveform based music composition and explore multi-persons dance choreography.

\newpage

\bibliographystyle{ACM-Reference-Format}
\balance
\bibliography{References}


\begin{thebibliography}{42}


\ifx \showCODEN    \undefined \def \showCODEN     #1{\unskip}     \fi
\ifx \showDOI      \undefined \def \showDOI       #1{#1}\fi
\ifx \showISBNx    \undefined \def \showISBNx     #1{\unskip}     \fi
\ifx \showISBNxiii \undefined \def \showISBNxiii  #1{\unskip}     \fi
\ifx \showISSN     \undefined \def \showISSN      #1{\unskip}     \fi
\ifx \showLCCN     \undefined \def \showLCCN      #1{\unskip}     \fi
\ifx \shownote     \undefined \def \shownote      #1{#1}          \fi
\ifx \showarticletitle \undefined \def \showarticletitle #1{#1}   \fi
\ifx \showURL      \undefined \def \showURL       {\relax}        \fi
\providecommand\bibfield[2]{#2}
\providecommand\bibinfo[2]{#2}
\providecommand\natexlab[1]{#1}
\providecommand\showeprint[2][]{arXiv:#2}

\bibitem[\protect\citeauthoryear{Alvarez-Melis and Jaakkola}{Alvarez-Melis and
  Jaakkola}{2018}]%
        {alvarez2018gromov}
\bibfield{author}{\bibinfo{person}{David Alvarez-Melis} {and}
  \bibinfo{person}{Tommi Jaakkola}.} \bibinfo{year}{2018}\natexlab{}.
\newblock \showarticletitle{Gromov-Wasserstein Alignment of Word Embedding
  Spaces}. In \bibinfo{booktitle}{\emph{Proceedings of the 2018 Conference on
  Empirical Methods in Natural Language Processing}}.
  \bibinfo{pages}{1881--1890}.
\newblock


\bibitem[\protect\citeauthoryear{Astey et~al\mbox{.}}{Astey
  et~al\mbox{.}}{1987}]%
        {astey1987cobordism}
\bibfield{author}{\bibinfo{person}{Luis Astey} {et~al\mbox{.}}}
  \bibinfo{year}{1987}\natexlab{}.
\newblock \showarticletitle{A cobordism obstruction to embedding manifolds}.
\newblock \bibinfo{journal}{\emph{Illinois Journal of Mathematics}}
  \bibinfo{volume}{31}, \bibinfo{number}{2} (\bibinfo{year}{1987}),
  \bibinfo{pages}{344--350}.
\newblock


\bibitem[\protect\citeauthoryear{Bogo, Kanazawa, Lassner, Gehler, Romero, and
  Black}{Bogo et~al\mbox{.}}{2016}]%
        {bogo2016keep}
\bibfield{author}{\bibinfo{person}{Federica Bogo}, \bibinfo{person}{Angjoo
  Kanazawa}, \bibinfo{person}{Christoph Lassner}, \bibinfo{person}{Peter
  Gehler}, \bibinfo{person}{Javier Romero}, {and} \bibinfo{person}{Michael~J
  Black}.} \bibinfo{year}{2016}\natexlab{}.
\newblock \showarticletitle{Keep it SMPL: Automatic estimation of 3D human pose
  and shape from a single image}. In \bibinfo{booktitle}{\emph{European
  conference on computer vision}}. Springer, \bibinfo{pages}{561--578}.
\newblock


\bibitem[\protect\citeauthoryear{Bourgain}{Bourgain}{1985}]%
        {bourgain1985lipschitz}
\bibfield{author}{\bibinfo{person}{Jean Bourgain}.}
  \bibinfo{year}{1985}\natexlab{}.
\newblock \showarticletitle{On Lipschitz embedding of finite metric spaces in
  Hilbert space}.
\newblock \bibinfo{journal}{\emph{Israel Journal of Mathematics}}
  \bibinfo{volume}{52}, \bibinfo{number}{1-2} (\bibinfo{year}{1985}),
  \bibinfo{pages}{46--52}.
\newblock


\bibitem[\protect\citeauthoryear{Briot, Hadjeres, and Pachet}{Briot
  et~al\mbox{.}}{2017}]%
        {briot2017deep}
\bibfield{author}{\bibinfo{person}{Jean-Pierre Briot},
  \bibinfo{person}{Ga{\"e}tan Hadjeres}, {and}
  \bibinfo{person}{Fran{\c{c}}ois-David Pachet}.}
  \bibinfo{year}{2017}\natexlab{}.
\newblock \showarticletitle{Deep learning techniques for music generation--a
  survey}.
\newblock \bibinfo{journal}{\emph{arXiv preprint arXiv:1709.01620}}
  (\bibinfo{year}{2017}).
\newblock


\bibitem[\protect\citeauthoryear{Brown and Parsons}{Brown and Parsons}{2008}]%
        {brown2008neuroscience}
\bibfield{author}{\bibinfo{person}{Steven Brown} {and}
  \bibinfo{person}{Lawrence~M Parsons}.} \bibinfo{year}{2008}\natexlab{}.
\newblock \showarticletitle{The neuroscience of dance}.
\newblock \bibinfo{journal}{\emph{Scientific American}} \bibinfo{volume}{299},
  \bibinfo{number}{1} (\bibinfo{year}{2008}), \bibinfo{pages}{78--83}.
\newblock


\bibitem[\protect\citeauthoryear{Child, Gray, Radford, and Sutskever}{Child
  et~al\mbox{.}}{2019}]%
        {child2019generating}
\bibfield{author}{\bibinfo{person}{Rewon Child}, \bibinfo{person}{Scott Gray},
  \bibinfo{person}{Alec Radford}, {and} \bibinfo{person}{Ilya Sutskever}.}
  \bibinfo{year}{2019}\natexlab{}.
\newblock \showarticletitle{Generating long sequences with sparse
  transformers}.
\newblock \bibinfo{journal}{\emph{arXiv preprint arXiv:1904.10509}}
  (\bibinfo{year}{2019}).
\newblock


\bibitem[\protect\citeauthoryear{Devlin, Chang, Lee, and Toutanova}{Devlin
  et~al\mbox{.}}{2018}]%
        {devlin2018bert}
\bibfield{author}{\bibinfo{person}{Jacob Devlin}, \bibinfo{person}{Ming-Wei
  Chang}, \bibinfo{person}{Kenton Lee}, {and} \bibinfo{person}{Kristina
  Toutanova}.} \bibinfo{year}{2018}\natexlab{}.
\newblock \showarticletitle{Bert: Pre-training of deep bidirectional
  transformers for language understanding}.
\newblock \bibinfo{journal}{\emph{arXiv preprint arXiv:1810.04805}}
  (\bibinfo{year}{2018}).
\newblock


\bibitem[\protect\citeauthoryear{Dhariwal, Jun, Payne, Kim, Radford, and
  Sutskever}{Dhariwal et~al\mbox{.}}{2020}]%
        {dhariwal2020jukebox}
\bibfield{author}{\bibinfo{person}{Prafulla Dhariwal}, \bibinfo{person}{Heewoo
  Jun}, \bibinfo{person}{Christine Payne}, \bibinfo{person}{Jong~Wook Kim},
  \bibinfo{person}{Alec Radford}, {and} \bibinfo{person}{Ilya Sutskever}.}
  \bibinfo{year}{2020}\natexlab{}.
\newblock \showarticletitle{Jukebox: A generative model for music}.
\newblock \bibinfo{journal}{\emph{arXiv preprint arXiv:2005.00341}}
  (\bibinfo{year}{2020}).
\newblock


\bibitem[\protect\citeauthoryear{Dieleman, Oord, and Simonyan}{Dieleman
  et~al\mbox{.}}{2018}]%
        {dieleman2018challenge}
\bibfield{author}{\bibinfo{person}{Sander Dieleman}, \bibinfo{person}{A{\"a}ron
  van~den Oord}, {and} \bibinfo{person}{Karen Simonyan}.}
  \bibinfo{year}{2018}\natexlab{}.
\newblock \showarticletitle{The challenge of realistic music generation:
  modelling raw audio at scale}.
\newblock \bibinfo{journal}{\emph{arXiv preprint arXiv:1806.10474}}
  (\bibinfo{year}{2018}).
\newblock


\bibitem[\protect\citeauthoryear{Fan, Xu, and Geng}{Fan et~al\mbox{.}}{2011}]%
        {fan2011example}
\bibfield{author}{\bibinfo{person}{Rukun Fan}, \bibinfo{person}{Songhua Xu},
  {and} \bibinfo{person}{Weidong Geng}.} \bibinfo{year}{2011}\natexlab{}.
\newblock \showarticletitle{Example-based automatic music-driven conventional
  dance motion synthesis}.
\newblock \bibinfo{journal}{\emph{IEEE transactions on visualization and
  computer graphics}} \bibinfo{volume}{18}, \bibinfo{number}{3}
  (\bibinfo{year}{2011}), \bibinfo{pages}{501--515}.
\newblock


\bibitem[\protect\citeauthoryear{Grifski}{Grifski}{[n.d.]}]%
        {JuxtaMidi}
\bibfield{author}{\bibinfo{person}{Jeremy Grifski}.}
  \bibinfo{year}{[n.d.]}\natexlab{}.
\newblock \bibinfo{title}{JuxtaMidi}.
\newblock
  \bibinfo{howpublished}{\url{https://therenegadecoder.com/code/juxtamidi-a-midi-file-visualization-dashboard/}}.
\newblock
\newblock
\shownote{Accessed: 2021-04-17.}


\bibitem[\protect\citeauthoryear{He, Xia, Qin, Wang, Yu, Liu, and Ma}{He
  et~al\mbox{.}}{2016}]%
        {he2016dual}
\bibfield{author}{\bibinfo{person}{Di He}, \bibinfo{person}{Yingce Xia},
  \bibinfo{person}{Tao Qin}, \bibinfo{person}{Liwei Wang},
  \bibinfo{person}{Nenghai Yu}, \bibinfo{person}{Tie-Yan Liu}, {and}
  \bibinfo{person}{Wei-Ying Ma}.} \bibinfo{year}{2016}\natexlab{}.
\newblock \showarticletitle{Dual learning for machine translation}.
\newblock \bibinfo{journal}{\emph{Advances in neural information processing
  systems}}  \bibinfo{volume}{29} (\bibinfo{year}{2016}),
  \bibinfo{pages}{820--828}.
\newblock


\bibitem[\protect\citeauthoryear{Huang, Vaswani, Uszkoreit, Shazeer, Simon,
  Hawthorne, Dai, Hoffman, Dinculescu, and Eck}{Huang et~al\mbox{.}}{2018}]%
        {huang2018music}
\bibfield{author}{\bibinfo{person}{Cheng-Zhi~Anna Huang},
  \bibinfo{person}{Ashish Vaswani}, \bibinfo{person}{Jakob Uszkoreit},
  \bibinfo{person}{Noam Shazeer}, \bibinfo{person}{Ian Simon},
  \bibinfo{person}{Curtis Hawthorne}, \bibinfo{person}{Andrew~M Dai},
  \bibinfo{person}{Matthew~D Hoffman}, \bibinfo{person}{Monica Dinculescu},
  {and} \bibinfo{person}{Douglas Eck}.} \bibinfo{year}{2018}\natexlab{}.
\newblock \showarticletitle{Music transformer}.
\newblock \bibinfo{journal}{\emph{arXiv preprint arXiv:1809.04281}}
  (\bibinfo{year}{2018}).
\newblock


\bibitem[\protect\citeauthoryear{Huang, Hu, Wu, Sawada, Zhang, and Jiang}{Huang
  et~al\mbox{.}}{2021}]%
        {huang2021dance}
\bibfield{author}{\bibinfo{person}{Ruozi Huang}, \bibinfo{person}{Huang Hu},
  \bibinfo{person}{Wei Wu}, \bibinfo{person}{Kei Sawada}, \bibinfo{person}{Mi
  Zhang}, {and} \bibinfo{person}{Daxin Jiang}.}
  \bibinfo{year}{2021}\natexlab{}.
\newblock \showarticletitle{Dance Revolution: Long-Term Dance Generation with
  Music via Curriculum Learning}. In \bibinfo{booktitle}{\emph{International
  Conference on Learning Representations}}.
\newblock


\bibitem[\protect\citeauthoryear{Huang and Yang}{Huang and Yang}{2020}]%
        {huang2020pop}
\bibfield{author}{\bibinfo{person}{Yu-Siang Huang} {and}
  \bibinfo{person}{Yi-Hsuan Yang}.} \bibinfo{year}{2020}\natexlab{}.
\newblock \showarticletitle{Pop Music Transformer: Beat-based modeling and
  generation of expressive Pop piano compositions}. In
  \bibinfo{booktitle}{\emph{Proceedings of the 28th ACM International
  Conference on Multimedia}}. \bibinfo{pages}{1180--1188}.
\newblock


\bibitem[\protect\citeauthoryear{Isola, Zhu, Zhou, and Efros}{Isola
  et~al\mbox{.}}{2017}]%
        {isola2017image}
\bibfield{author}{\bibinfo{person}{Phillip Isola}, \bibinfo{person}{Jun-Yan
  Zhu}, \bibinfo{person}{Tinghui Zhou}, {and} \bibinfo{person}{Alexei~A
  Efros}.} \bibinfo{year}{2017}\natexlab{}.
\newblock \showarticletitle{Image-to-image translation with conditional
  adversarial networks}. In \bibinfo{booktitle}{\emph{Proceedings of the IEEE
  conference on computer vision and pattern recognition}}.
  \bibinfo{pages}{1125--1134}.
\newblock


\bibitem[\protect\citeauthoryear{Lee, Yang, Liu, Wang, Lu, Yang, and Kautz}{Lee
  et~al\mbox{.}}{2019}]%
        {lee2019dancing}
\bibfield{author}{\bibinfo{person}{Hsin-Ying Lee}, \bibinfo{person}{Xiaodong
  Yang}, \bibinfo{person}{Ming-Yu Liu}, \bibinfo{person}{Ting-Chun Wang},
  \bibinfo{person}{Yu-Ding Lu}, \bibinfo{person}{Ming-Hsuan Yang}, {and}
  \bibinfo{person}{Jan Kautz}.} \bibinfo{year}{2019}\natexlab{}.
\newblock \showarticletitle{Dancing to music}. In
  \bibinfo{booktitle}{\emph{Advances in Neural Information Processing
  Systems}}. \bibinfo{pages}{3586--3596}.
\newblock


\bibitem[\protect\citeauthoryear{Lee, Lee, and Park}{Lee et~al\mbox{.}}{2013}]%
        {lee2013music}
\bibfield{author}{\bibinfo{person}{Minho Lee}, \bibinfo{person}{Kyogu Lee},
  {and} \bibinfo{person}{Jaeheung Park}.} \bibinfo{year}{2013}\natexlab{}.
\newblock \showarticletitle{Music similarity-based approach to generating dance
  motion sequence}.
\newblock \bibinfo{journal}{\emph{Multimedia tools and applications}}
  \bibinfo{volume}{62}, \bibinfo{number}{3} (\bibinfo{year}{2013}),
  \bibinfo{pages}{895--912}.
\newblock


\bibitem[\protect\citeauthoryear{Levitin and Tirovolas}{Levitin and
  Tirovolas}{2009}]%
        {levitin2009current}
\bibfield{author}{\bibinfo{person}{Daniel~J Levitin} {and}
  \bibinfo{person}{Anna~K Tirovolas}.} \bibinfo{year}{2009}\natexlab{}.
\newblock \showarticletitle{Current advances in the cognitive neuroscience of
  music}.
\newblock \bibinfo{journal}{\emph{Annals of the New York Academy of Sciences}}
  \bibinfo{volume}{1156}, \bibinfo{number}{1} (\bibinfo{year}{2009}),
  \bibinfo{pages}{211--231}.
\newblock


\bibitem[\protect\citeauthoryear{Li, Yin, Chu, Zhou, Wang, Fidler, and Li}{Li
  et~al\mbox{.}}{2020}]%
        {li2020learning}
\bibfield{author}{\bibinfo{person}{Jiaman Li}, \bibinfo{person}{Yihang Yin},
  \bibinfo{person}{Hang Chu}, \bibinfo{person}{Yi Zhou},
  \bibinfo{person}{Tingwu Wang}, \bibinfo{person}{Sanja Fidler}, {and}
  \bibinfo{person}{Hao Li}.} \bibinfo{year}{2020}\natexlab{}.
\newblock \showarticletitle{Learning to Generate Diverse Dance Motions with
  Transformer}.
\newblock \bibinfo{journal}{\emph{arXiv preprint arXiv:2008.08171}}
  (\bibinfo{year}{2020}).
\newblock


\bibitem[\protect\citeauthoryear{Li, Yang, Ross, and Kanazawa}{Li
  et~al\mbox{.}}{2021}]%
        {li2021learn}
\bibfield{author}{\bibinfo{person}{Ruilong Li}, \bibinfo{person}{Shan Yang},
  \bibinfo{person}{David~A Ross}, {and} \bibinfo{person}{Angjoo Kanazawa}.}
  \bibinfo{year}{2021}\natexlab{}.
\newblock \showarticletitle{Learn to Dance with AIST++: Music Conditioned 3D
  Dance Generation}.
\newblock \bibinfo{journal}{\emph{arXiv preprint arXiv:2101.08779}}
  (\bibinfo{year}{2021}).
\newblock


\bibitem[\protect\citeauthoryear{Lin, Xia, Qin, Chen, and Liu}{Lin
  et~al\mbox{.}}{2018}]%
        {lin2018conditional}
\bibfield{author}{\bibinfo{person}{Jianxin Lin}, \bibinfo{person}{Yingce Xia},
  \bibinfo{person}{Tao Qin}, \bibinfo{person}{Zhibo Chen}, {and}
  \bibinfo{person}{Tie-Yan Liu}.} \bibinfo{year}{2018}\natexlab{}.
\newblock \showarticletitle{Conditional image-to-image translation}. In
  \bibinfo{booktitle}{\emph{Proceedings of the IEEE conference on computer
  vision and pattern recognition}}. \bibinfo{pages}{5524--5532}.
\newblock


\bibitem[\protect\citeauthoryear{Loper, Mahmood, Romero, Pons-Moll, and
  Black}{Loper et~al\mbox{.}}{2015}]%
        {loper2015smpl}
\bibfield{author}{\bibinfo{person}{Matthew Loper}, \bibinfo{person}{Naureen
  Mahmood}, \bibinfo{person}{Javier Romero}, \bibinfo{person}{Gerard
  Pons-Moll}, {and} \bibinfo{person}{Michael~J Black}.}
  \bibinfo{year}{2015}\natexlab{}.
\newblock \showarticletitle{SMPL: A skinned multi-person linear model}.
\newblock \bibinfo{journal}{\emph{ACM transactions on graphics (TOG)}}
  \bibinfo{volume}{34}, \bibinfo{number}{6} (\bibinfo{year}{2015}),
  \bibinfo{pages}{1--16}.
\newblock


\bibitem[\protect\citeauthoryear{McFee, Raffel, Liang, Ellis, McVicar,
  Battenberg, and Nieto}{McFee et~al\mbox{.}}{2015}]%
        {mcfee2015librosa}
\bibfield{author}{\bibinfo{person}{Brian McFee}, \bibinfo{person}{Colin
  Raffel}, \bibinfo{person}{Dawen Liang}, \bibinfo{person}{Daniel~PW Ellis},
  \bibinfo{person}{Matt McVicar}, \bibinfo{person}{Eric Battenberg}, {and}
  \bibinfo{person}{Oriol Nieto}.} \bibinfo{year}{2015}\natexlab{}.
\newblock \showarticletitle{librosa: Audio and music signal analysis in
  python}. In \bibinfo{booktitle}{\emph{Proceedings of the 14th python in
  science conference}}, Vol.~\bibinfo{volume}{8}. \bibinfo{pages}{18--25}.
\newblock


\bibitem[\protect\citeauthoryear{M{\'e}moli}{M{\'e}moli}{2011}]%
        {memoli2011gromov}
\bibfield{author}{\bibinfo{person}{Facundo M{\'e}moli}.}
  \bibinfo{year}{2011}\natexlab{}.
\newblock \showarticletitle{Gromov--Wasserstein distances and the metric
  approach to object matching}.
\newblock \bibinfo{journal}{\emph{Foundations of computational mathematics}}
  \bibinfo{volume}{11}, \bibinfo{number}{4} (\bibinfo{year}{2011}),
  \bibinfo{pages}{417--487}.
\newblock


\bibitem[\protect\citeauthoryear{M{\"u}ller}{M{\"u}ller}{2015}]%
        {muller2015fundamentals}
\bibfield{author}{\bibinfo{person}{Meinard M{\"u}ller}.}
  \bibinfo{year}{2015}\natexlab{}.
\newblock \bibinfo{booktitle}{\emph{Fundamentals of music processing: Audio,
  analysis, algorithms, applications}}.
\newblock \bibinfo{publisher}{Springer}.
\newblock


\bibitem[\protect\citeauthoryear{Peyr{\'e}, Cuturi, and Solomon}{Peyr{\'e}
  et~al\mbox{.}}{2016}]%
        {peyre2016gromov}
\bibfield{author}{\bibinfo{person}{Gabriel Peyr{\'e}}, \bibinfo{person}{Marco
  Cuturi}, {and} \bibinfo{person}{Justin Solomon}.}
  \bibinfo{year}{2016}\natexlab{}.
\newblock \showarticletitle{Gromov-wasserstein averaging of kernel and distance
  matrices}. In \bibinfo{booktitle}{\emph{International Conference on Machine
  Learning}}. \bibinfo{pages}{2664--2672}.
\newblock


\bibitem[\protect\citeauthoryear{Ren, Li, Huang, and Chen}{Ren
  et~al\mbox{.}}{2020}]%
        {ren2020self}
\bibfield{author}{\bibinfo{person}{Xuanchi Ren}, \bibinfo{person}{Haoran Li},
  \bibinfo{person}{Zijian Huang}, {and} \bibinfo{person}{Qifeng Chen}.}
  \bibinfo{year}{2020}\natexlab{}.
\newblock \showarticletitle{Self-supervised Dance Video Synthesis Conditioned
  on Music}. In \bibinfo{booktitle}{\emph{Proceedings of the 28th ACM
  International Conference on Multimedia}}. \bibinfo{pages}{46--54}.
\newblock


\bibitem[\protect\citeauthoryear{Small}{Small}{1998}]%
        {small1998musicking}
\bibfield{author}{\bibinfo{person}{Christopher Small}.}
  \bibinfo{year}{1998}\natexlab{}.
\newblock \bibinfo{booktitle}{\emph{Musicking: The meanings of performing and
  listening}}.
\newblock \bibinfo{publisher}{Wesleyan University Press}.
\newblock


\bibitem[\protect\citeauthoryear{Tang, Jia, and Mao}{Tang
  et~al\mbox{.}}{2018}]%
        {tang2018dance}
\bibfield{author}{\bibinfo{person}{Taoran Tang}, \bibinfo{person}{Jia Jia},
  {and} \bibinfo{person}{Hanyang Mao}.} \bibinfo{year}{2018}\natexlab{}.
\newblock \showarticletitle{Dance with melody: An LSTM-autoencoder approach to
  music-oriented dance synthesis}. In \bibinfo{booktitle}{\emph{Proceedings of
  the 26th ACM international conference on Multimedia}}.
  \bibinfo{pages}{1598--1606}.
\newblock


\bibitem[\protect\citeauthoryear{Tsuchida, Fukayama, Hamasaki, and
  Goto}{Tsuchida et~al\mbox{.}}{2019}]%
        {tsuchida2019aist}
\bibfield{author}{\bibinfo{person}{Shuhei Tsuchida}, \bibinfo{person}{Satoru
  Fukayama}, \bibinfo{person}{Masahiro Hamasaki}, {and}
  \bibinfo{person}{Masataka Goto}.} \bibinfo{year}{2019}\natexlab{}.
\newblock \showarticletitle{AIST Dance Video Database: Multi-Genre,
  Multi-Dancer, and Multi-Camera Database for Dance Information Processing.}.
  In \bibinfo{booktitle}{\emph{ISMIR}}. \bibinfo{pages}{501--510}.
\newblock


\bibitem[\protect\citeauthoryear{Vasquez and Lewis}{Vasquez and Lewis}{2019}]%
        {vasquez2019melnet}
\bibfield{author}{\bibinfo{person}{Sean Vasquez} {and} \bibinfo{person}{Mike
  Lewis}.} \bibinfo{year}{2019}\natexlab{}.
\newblock \showarticletitle{Melnet: A generative model for audio in the
  frequency domain}.
\newblock \bibinfo{journal}{\emph{arXiv preprint arXiv:1906.01083}}
  (\bibinfo{year}{2019}).
\newblock


\bibitem[\protect\citeauthoryear{Vaswani, Shazeer, Parmar, Uszkoreit, Jones,
  Gomez, Kaiser, and Polosukhin}{Vaswani et~al\mbox{.}}{2017}]%
        {vaswani2017attention}
\bibfield{author}{\bibinfo{person}{Ashish Vaswani}, \bibinfo{person}{Noam
  Shazeer}, \bibinfo{person}{Niki Parmar}, \bibinfo{person}{Jakob Uszkoreit},
  \bibinfo{person}{Llion Jones}, \bibinfo{person}{Aidan~N Gomez},
  \bibinfo{person}{Lukasz Kaiser}, {and} \bibinfo{person}{Illia Polosukhin}.}
  \bibinfo{year}{2017}\natexlab{}.
\newblock \showarticletitle{Attention is all you need}.
\newblock \bibinfo{journal}{\emph{arXiv preprint arXiv:1706.03762}}
  (\bibinfo{year}{2017}).
\newblock


\bibitem[\protect\citeauthoryear{Wang}{Wang}{2015}]%
        {wang2015hypothesis}
\bibfield{author}{\bibinfo{person}{Tianyan Wang}.}
  \bibinfo{year}{2015}\natexlab{}.
\newblock \showarticletitle{A hypothesis on the biological origins and social
  evolution of music and dance}.
\newblock \bibinfo{journal}{\emph{Frontiers in neuroscience}}
  \bibinfo{volume}{9} (\bibinfo{year}{2015}), \bibinfo{pages}{30}.
\newblock


\bibitem[\protect\citeauthoryear{Xia, Qin, Chen, Bian, Yu, and Liu}{Xia
  et~al\mbox{.}}{2017}]%
        {xia2017dual}
\bibfield{author}{\bibinfo{person}{Yingce Xia}, \bibinfo{person}{Tao Qin},
  \bibinfo{person}{Wei Chen}, \bibinfo{person}{Jiang Bian},
  \bibinfo{person}{Nenghai Yu}, {and} \bibinfo{person}{Tie-Yan Liu}.}
  \bibinfo{year}{2017}\natexlab{}.
\newblock \showarticletitle{Dual supervised learning}. In
  \bibinfo{booktitle}{\emph{International Conference on Machine Learning}}.
  PMLR, \bibinfo{pages}{3789--3798}.
\newblock


\bibitem[\protect\citeauthoryear{Xia, Tan, Tian, Qin, Yu, and Liu}{Xia
  et~al\mbox{.}}{2018}]%
        {xia2018model}
\bibfield{author}{\bibinfo{person}{Yingce Xia}, \bibinfo{person}{Xu Tan},
  \bibinfo{person}{Fei Tian}, \bibinfo{person}{Tao Qin},
  \bibinfo{person}{Nenghai Yu}, {and} \bibinfo{person}{Tie-Yan Liu}.}
  \bibinfo{year}{2018}\natexlab{}.
\newblock \showarticletitle{Model-level dual learning}. In
  \bibinfo{booktitle}{\emph{International Conference on Machine Learning}}.
  PMLR, \bibinfo{pages}{5383--5392}.
\newblock


\bibitem[\protect\citeauthoryear{Xu, Duan, Cai, Chia, Xu, and Tian}{Xu
  et~al\mbox{.}}{2004}]%
        {xu2004hmm}
\bibfield{author}{\bibinfo{person}{Min Xu}, \bibinfo{person}{Ling-Yu Duan},
  \bibinfo{person}{Jianfei Cai}, \bibinfo{person}{Liang-Tien Chia},
  \bibinfo{person}{Changsheng Xu}, {and} \bibinfo{person}{Qi Tian}.}
  \bibinfo{year}{2004}\natexlab{}.
\newblock \showarticletitle{HMM-based audio keyword generation}. In
  \bibinfo{booktitle}{\emph{Pacific-Rim Conference on Multimedia}}. Springer,
  \bibinfo{pages}{566--574}.
\newblock


\bibitem[\protect\citeauthoryear{Ye, Wu, Jia, Bu, Chen, Meng, and Wang}{Ye
  et~al\mbox{.}}{2020}]%
        {ye2020choreonet}
\bibfield{author}{\bibinfo{person}{Zijie Ye}, \bibinfo{person}{Haozhe Wu},
  \bibinfo{person}{Jia Jia}, \bibinfo{person}{Yaohua Bu}, \bibinfo{person}{Wei
  Chen}, \bibinfo{person}{Fanbo Meng}, {and} \bibinfo{person}{Yanfeng Wang}.}
  \bibinfo{year}{2020}\natexlab{}.
\newblock \showarticletitle{ChoreoNet: Towards Music to Dance Synthesis with
  Choreographic Action Unit}. In \bibinfo{booktitle}{\emph{Proceedings of the
  28th ACM International Conference on Multimedia}}. \bibinfo{pages}{744--752}.
\newblock


\bibitem[\protect\citeauthoryear{Yi, Zhang, Tan, and Gong}{Yi
  et~al\mbox{.}}{2017}]%
        {yi2017dualgan}
\bibfield{author}{\bibinfo{person}{Zili Yi}, \bibinfo{person}{Hao Zhang},
  \bibinfo{person}{Ping Tan}, {and} \bibinfo{person}{Minglun Gong}.}
  \bibinfo{year}{2017}\natexlab{}.
\newblock \showarticletitle{Dualgan: Unsupervised dual learning for
  image-to-image translation}. In \bibinfo{booktitle}{\emph{Proceedings of the
  IEEE international conference on computer vision}}.
  \bibinfo{pages}{2849--2857}.
\newblock


\bibitem[\protect\citeauthoryear{Zhou, Barnes, Lu, Yang, and Li}{Zhou
  et~al\mbox{.}}{2019}]%
        {zhou2019continuity}
\bibfield{author}{\bibinfo{person}{Yi Zhou}, \bibinfo{person}{Connelly Barnes},
  \bibinfo{person}{Jingwan Lu}, \bibinfo{person}{Jimei Yang}, {and}
  \bibinfo{person}{Hao Li}.} \bibinfo{year}{2019}\natexlab{}.
\newblock \showarticletitle{On the continuity of rotation representations in
  neural networks}. In \bibinfo{booktitle}{\emph{Proceedings of the IEEE
  Conference on Computer Vision and Pattern Recognition}}.
  \bibinfo{pages}{5745--5753}.
\newblock


\bibitem[\protect\citeauthoryear{Zhu, Park, Isola, and Efros}{Zhu
  et~al\mbox{.}}{2017}]%
        {zhu2017unpaired}
\bibfield{author}{\bibinfo{person}{Jun-Yan Zhu}, \bibinfo{person}{Taesung
  Park}, \bibinfo{person}{Phillip Isola}, {and} \bibinfo{person}{Alexei~A
  Efros}.} \bibinfo{year}{2017}\natexlab{}.
\newblock \showarticletitle{Unpaired image-to-image translation using
  cycle-consistent adversarial networks}. In
  \bibinfo{booktitle}{\emph{Proceedings of the IEEE international conference on
  computer vision}}. \bibinfo{pages}{2223--2232}.
\newblock


\end{thebibliography}
\end{document}